\documentclass[twocolumn]{aastex631}

\submitjournal{PSJ}

\shorttitle{polar vortices on terrestrial planets}
\shortauthors{Guendelman et al.}
\begin{document}

\title{Dynamical regimes of polar vortices on terrestrial planets with a seasonal cycle}

\correspondingauthor{Ilai Guendelman}
\email{ilai.guendelman@weizmann.ac.il}

\author[0000-0002-6873-0320]{Ilai Guendelman}
\affiliation{Department of Earth and Planetary Sciences, Weizmann Institute of \
Science \\
234 Herzl st., 76100 \\
Rehovot, Israel}

\author[0000-0003-4089-0020]{Darryn W. Waugh}
\affiliation{Department of Earth and Planetary Sciences, Johns Hopkins University\\
3400 North Charles Street, 21218 \\
Baltimore, Maryland}

\author[0000-0003-4089-0020]{Yohai Kaspi}
\affiliation{Department of Earth and Planetary Sciences, Weizmann Institute of \
Science \\
234 Herzl st., 76100 \\
Rehovot, Israel}



\begin{abstract}
Polar vortices are common planetary flows that encircle the pole in the middle or high latitudes, and are observed on most of the solar systems' planetary atmospheres. The polar vortices on Earth, Mars, and Titan are dynamically related to the mean meridional circulation and exhibits a significant seasonal cycle. However, the polar vortex's characteristics vary between the three planets. To understand the mechanisms that influence the polar vortex's dynamics and dependence on planetary parameters, we use an idealized general circulation model with a seasonal cycle in which we varied the obliquity, rotation rate, and orbital period. We find that there are distinct regimes for the polar vortex seasonal cycle across the parameter space. Some regimes have similarities to the observed polar vortices, including a weakening of the polar vortex during midwinter at slow rotation rates, similar to Titan's polar vortex. However, other regimes found within the parameter space have no counterpart in the solar system. In addition, we show that for a significant fraction of the parameter space, the vortex's potential vorticity latitudinal structure is annular, similar to the observed structure of the polar vortex on Mars and Titan. We also find a suppression of storm activity during midwinter that resembles the suppression observed on Mars and Earth, which occurs in simulations where the jet speed is greater than ~60 ms$^{-1}$. This wide variety of polar vortex dynamical regimes that shares similarities to observed polar vortices suggests that among exoplanets, there can be a wide variability of polar vortices.
\end{abstract}


\section{Introduction}
% \the\textwidth
% \\
% \the\columnwidth\\

\begin{figure*}[htb!]
    \centering
    \includegraphics[width=16cm]{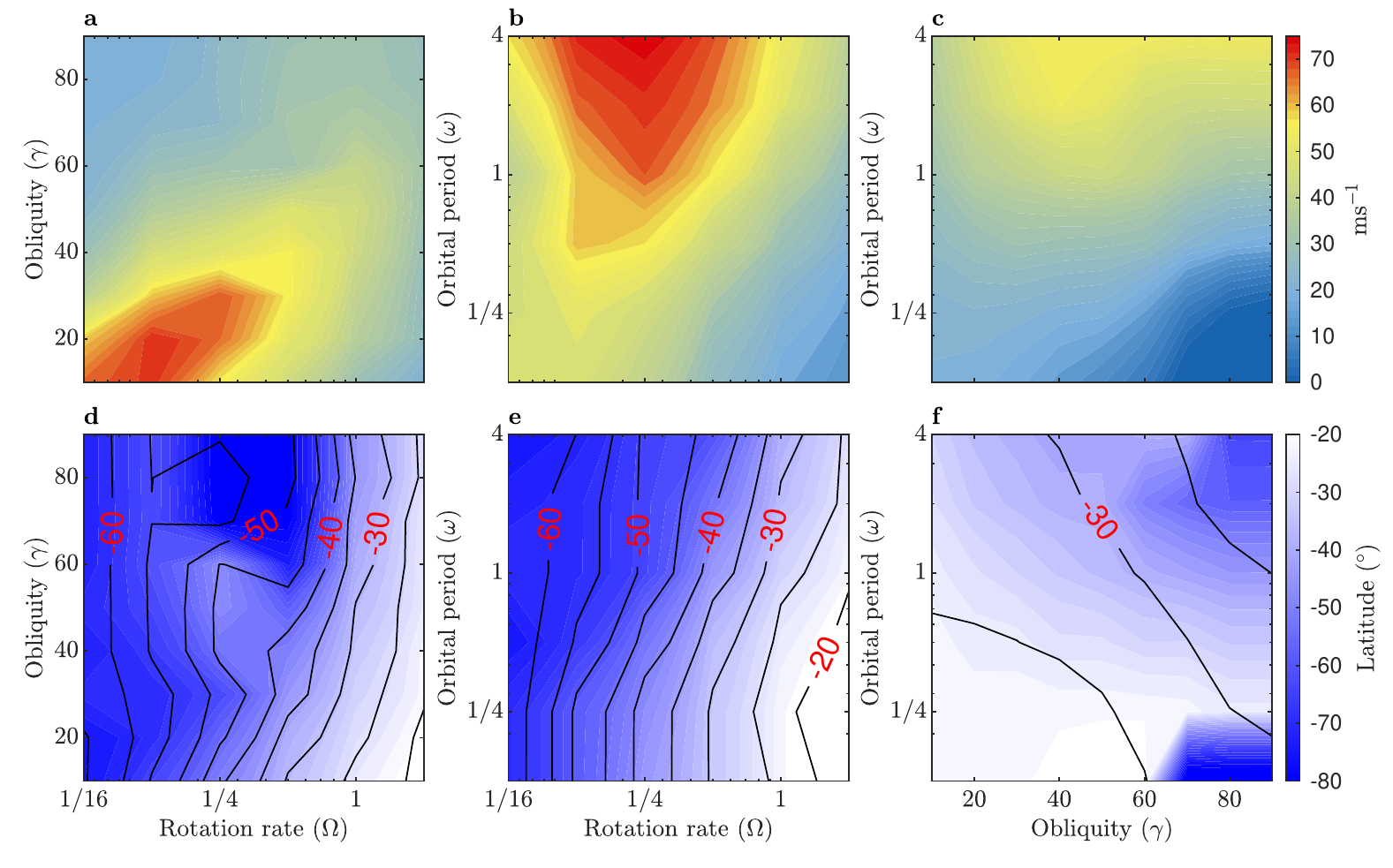}
    \caption{Top row: The maximum zonal mean zonal wind speed (ms$^{-1}$), for a vertical mean between $\sigma=0.25$ and $\sigma=0.5$ in the southern hemisphere during southern hemisphere winter. Bottom row: The latitude of maximum winds (shading) and latitude of descending branch of the Hadley cell. In a and d, $\omega=1$, in b and e, $\gamma=30^{\circ}$, and in c and f, $\Omega=1$.}
    \label{fig:fig1}
\end{figure*}

Polar vortices are a ubiquitous feature of planetary atmospheres, with Earth, Venus, Mars, Titan, Jupiter, Saturn, and possibly Neptune and Pluto all exhibiting polar vortices \citep[e.g.,][]{french_martian_1979,teanby_titans_2008,dyudina_dynamics_2008,luz_venuss_2011,polvani_stratosphere_2010,mitchell_polar_2015, Adriani2018, Gavriel2021, mitchell_2021}. These polar vortices are a fundamental aspect of the atmospheric circulation, and can be the site of unique microphysical and chemical processes, e.g., ozone depletion in Earth’s stratosphere, CO$_2$ condensation on Mars, and HCN clouds on Titan. There is no unique definition of a polar vortex, but here we follow \citet{waugh_what_2017} and define polar vortices as strong, planetary-scale flows that encircle the pole in the middle or high latitudes. The edge of such vortex can be defined by the latitude at which the zonal wind reaches its hemispheric maximum, which is also referred to as the latitude of the atmospheric jet. Alternatively, it can be defined by a region of high potential vorticity (PV), with the vortex edge located at the latitude with the steepest meridional PV gradients. Here we use both the terms polar vortex and jet, with polar vortex used primarily when discussing the PV structure, and jet when discussing the zonal wind structure.

While polar vortices are common, their characteristics vary among the solar system planets. This includes not only variations in the size and strength of the polar vortices, but also in their seasonal variability. For example, the polar vortices on Earth and Mars are generally strongest (strongest winds and steepest PV gradients) around or after the winter solstice \citep[e.g.,][]{mitchell_polar_2015, waugh_what_2017}, whereas observations and model simulations suggest that Titan's polar vortex is strongest during late fall and weakens during midwinter \citep[e.g.,][]{lora_gcm_2015, teanby_formation_2017, teanby_seasonal_2019, shultis_2022}. The polar vortices on these planets also differ in their PV structure: On Earth, the PV generally has a monopolar meridional structure, with PV maximizing (in absolute terms) at or near the pole \citep[e.g.,][]{polvani_stratosphere_2010}, whereas on Mars and Titan, there is an annular structure with the maximum PV located away from the pole \citep[e.g.,][]{mitchell_polar_2015, waugh_martian_2016,sharkey_mapping_2020,sharkey_potential_2021}. The causes of some of the differences are known, but others are not explained and there is a need to better understand the underlying mechanisms controlling the polar vortex dynamics on the different planets. This will not only improve the understanding of atmospheric dynamics on these planets, but will also provide insights into the structure/evolution of polar vortices on terrestrial exoplanets. 

Another topic where a better understanding is needed is the relation between polar vortices and storm activity. In Earth’s troposphere, storm activity over the North Pacific ocean weakens during midwinter even though the jet is strongest during this period \citep{nakamura_midwinter_1992}. This phenomenon also appears over the North Atlantic ocean during years with a strong jet \citep{afargan_midwinter_2017}. The suppression of storm activity during midwinter is in odds with linear models of baroclinic instability, but can be explained when considering the dephasing of the baroclinic wave when the jet is stronger \citep{hadas_2020}. On Mars, similar to Earth, there is also a suppression of baroclinic activity during solstice, which is again the period when the jet is strongest \citep[e.g.,][]{LEWIS2016, MULHOLLAND2016, LEE2018}. Although there are similarities between the two planets, there are also some differences in the characteristics of the suppression that occurs on both planets \citep{LEWIS2016}. The fact that on both planets there is a suppression of storm activity during the period with the strongest jet suggests that this suppression can occur in other planets and there is a need to better understand how common this phenomenon is in planetary atmospheres. 

In this study, we seek to understand the dynamics governing these phenomena by varying three planetary parameters, the rotation rate, obliquity, and orbital period, in an idealized three-dimensional general circulation model (GCM). Using this large suite of simulations, we analyze the polar vortex structure, seasonal evolution and seasonal relation to storm activity. Although the idealized model neglects some processes, such as CO$_2$ condensation, which was shown to be essential for the polar vortex characteristics on Mars \citep[e.g.,][]{toigo_what_2017,seviour_stability_2017}, it simplifies the problem so as to allow the identification of the key parameters and dynamical processes.

The model used in this study is described in section \ref{sec:model}. In section \ref{sec:strength}, we examine the dependence of the jet strength and seasonality on the planetary parameters. We show that the flows observed on the solar system terrestrial planets are found within our parameter space and align with the respective planet's parameters. Additionally, we show new regimes within the parameter space, that do not resemble the flow observed in the solar system terrestrial atmospheres. In section \ref{sec:EKE}, we discuss the relationship between the jet strength and storm activity; we show that a midwinter minimum in storm activity occurs in our parameter space. Following that, we discuss the PV structure of the polar vortex in section \ref{sec:PV}, showing that a state of an annular PV, similar to that observed on Mars and Titan, is common. We conclude in section \ref{sec:conc}.

\section{Model and simulations}\label{sec:model}

\begin{figure*}[htb!]
    \centering
    \includegraphics[width=16cm]{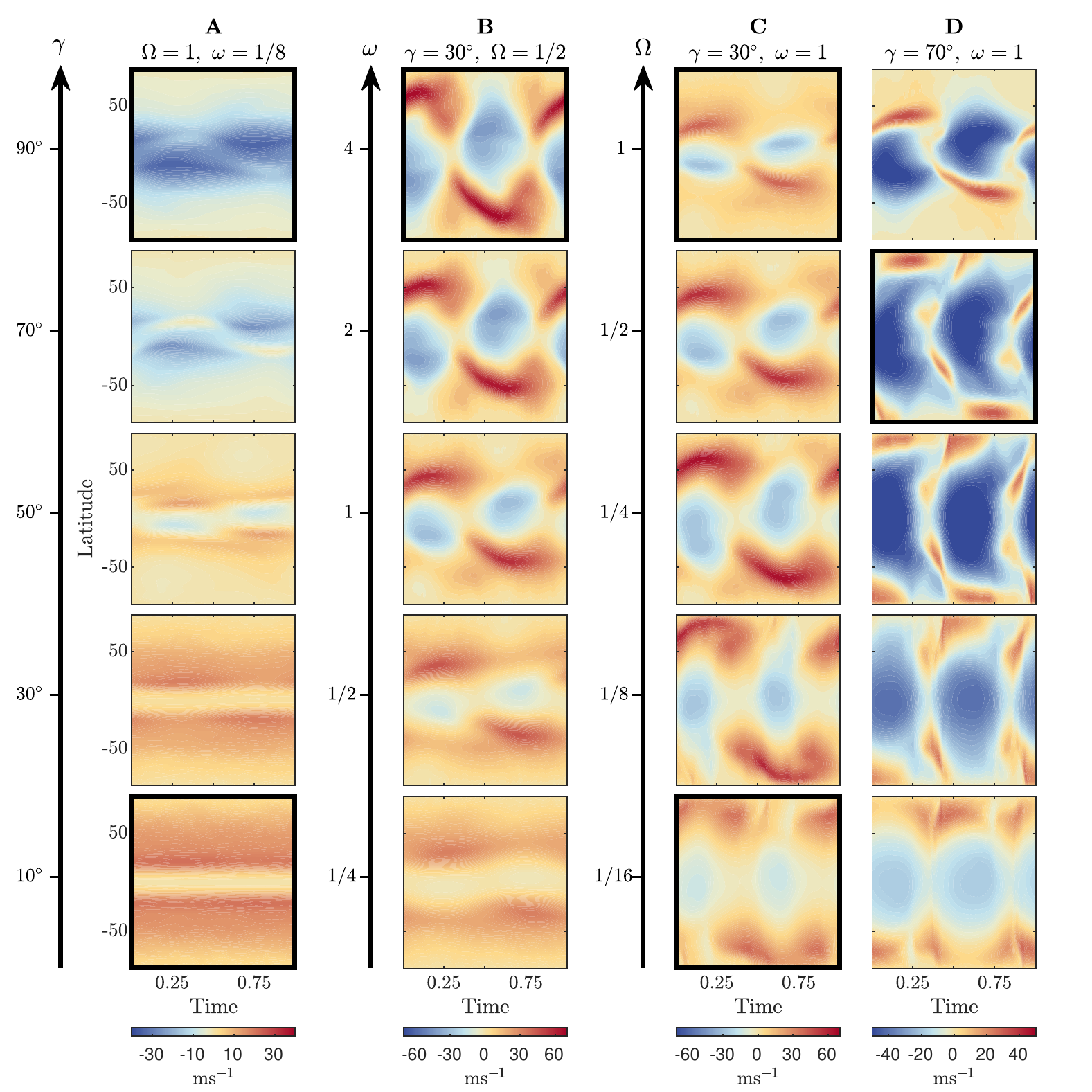}
    \caption{Selected simulations demonstrating the most dominant types of seasonal behavior within the parameter space. Plots show the zonal mean zonal wind vertically averaged between $\sigma=0.25$ and $\sigma=0.5$. The x-axis represents the year fraction; the year starts during the southern summer solstice. Column A shows the jet dependence on obliquity ($\gamma$) in the case of a short orbital period ($\omega = 1/8$) and an Earth-like rotation rate ($\Omega=1$), i.e., weak seasonality. Column B shows the jet's dependence on the orbital period ($\omega$) for a moderate obliquity ($\gamma=30^{\circ}$) and rotation rate ($\Omega=1/2$). Columns C and D show the jet's dependence on the rotation rate ($\Omega$) for an Earth-like orbital period ($\omega = 1$), and an obliquity of $30^{\circ}$ and $70^{\circ}$, respectively. Highlighted panels are the representative cases of the distinct jet behavior in the parameter space and are summarized in Figure~\ref{fig:fig3}. Note that there is a different colorscale for each column.}
    \label{fig:fig2}
\end{figure*}

To understand the variability of the polar vortex's seasonal cycle in planetary atmospheres, we use an idealized GCM with a seasonal cycle based on the GFDL dynamical core \citep{guendelman_atmospheric_2019}. The model is an aquaplanet GCM with a $10$m depth slab ocean. The model is forced at the top of the atmosphere with a diurnal mean seasonal insolation. The radiation transfer in the atmosphere is represented using a two-stream radiation scheme \citep{frierson2006} where we keep the optical depth constant in latitude. The model utilizes a simplified parameterization for moist convection \citep{frierson2006} and neglects different effects such as clouds and ice. The simplicity of the model's radiation scheme and the lack of an ozone layer results in a different flow in the model's stratosphere compared to that of Earth \citep[e.g.,][]{tan2019}. 

We run simulations in which we vary the rotation rate ($\Omega$) from 1/16 to 2 times Earth's rotation rate, the obliquity ($\gamma$) from $10^{\circ}$ to $90^{\circ}$, and the orbital period ($\omega$) from 1/8 to 4 times Earth's orbital period, using $360$ days in an Earth-like orbital period (for example, northern hemisphere solstice is in the middle of the year). We run three different sub-spaces of this parameter space by varying two parameters and keeping one constant (the constant values are $\gamma=30^{\circ}, \ \Omega=1$ and $\omega=1$). We run the simulations with a T42 resolution and 25 vertical $\sigma$ levels ($\sigma=p/p_s$, where $p_s$ is the surface pressure). All simulations were run for at least $80$ Earth years ($360$ days), and the climatology is calculated using the latter $50$ years (based on the simulation's orbital period). 

The range for the parameters is chosen to cover a large variety of climates. For the obliquity, we cover the entire range. We use the solar system planets as a guide for the rotation rate, taking the lower and upper edge to be close to Titan's and Jupiter's rotation rates, respectively. The upper edge of the orbital period was taken to mitigate the computational cost, as longer orbital periods requires longer computations without an increase in benefit for the study. Both the lower edge of the rotation rate and orbital period are taken such that the use of diurnal mean forcing is justified \citep{Salameh2018, guendelman_atmospheric_2019}, i.e., we do not include tidally locked planets in this study. Note, that although the model uses an Earth-like configuration, e.g., has moist convection with water as a condensible and an ocean mixed layer as a boundary, the model's simplicity allows us to draw conclusions regarding other planets as well.

Each of the three parameters has a different effect on the planetary climate. The degree of seasonality is controlled mainly by the planet's obliquity and orbital period. The obliquity resolves the latitudinal seasonal cycle of the top of the atmosphere solar forcing. The orbital period controls the timescale that the atmosphere has to adjust to radiative changes and plays a crucial role in the resulting seasonal cycle. Unlike the previous two parameters, which relate mainly to the atmosphere's radiative forcing, the rotation rate is a crucial parameter that strongly affects the atmospheric circulation \citep{kaspi_atmospheric_2015}.  

\section{Jet strength and seasonality}\label{sec:strength}
\subsection{Parameter space overview}

Jets persist across the parameter space explored. In most cases, the strongest jet is in the winter hemisphere. However, for strong seasonality and slow rotation rates there are weak winds in the winter hemisphere and the prominent vortex is in the summer hemisphere \citep{guendelman_2021}. In this study, we will mainly focus on the winter jet.

The strength and latitude of the jet varies significantly within the explored parameter space. Figure~\ref{fig:fig1} shows the dependence of the maximum wind speed (top row) and the latitude of the winter jet (bottom row, colors) on the different parameters. The wind speed changes non-monotonically with the rotation rate (Fig.~\ref{fig:fig1}a-b); a similar trend was noticed and explained using angular momentum conservation arguments in previous studies \citep{kaspi_atmospheric_2015, wang2018, guendelman_2021}. The wind strength's dependence on the orbital period is monotonic, where the wind strength increases for longer orbital periods (Fig.~\ref{fig:fig1}b-c). The dependence of the maximum wind speed on obliquity depends on the orbital period. For long orbital periods, the dependence is non-monotonic, with the wind speed maximizing at moderate obliquity values (Fig.~\ref{fig:fig1}a,c), which can be explained using considerations of angular momentum conservation \citep{guendelman_2021}. For short orbital periods, the maximum wind speed decreases monotonically with increasing obliquity (Fig.~\ref{fig:fig1}c). This monotonic decrease with obliquity is a result of the dominance of the annual mean climate at short orbital periods. As the obliquity increases, the annual mean insolation meridional gradient decreases, and reverses for obliquities larger than $54^{\circ}$ \citep{guendelman_atmospheric_2019}. This, in turn, results in a decrease in the temperature meridional gradient and a weakening of the jet.

The variation in the jet's latitude is in good agreement with the latitude of the descending edge of the Hadley circulation (Fig.~\ref{fig:fig1}d-e), indicating that the jet correlates to the mean meridional circulation. That said, there is a misfit for short orbital periods and high obliquity (Fig.~\ref{fig:fig1}f), where there are weak westerlies and the maximum wind speed is close to zero (Fig.~\ref{fig:fig1}c), i.e., there is no winter polar vortex. In cases of high obliquities and short orbital periods, the climate has a weak seasonal cycle and reversed temperature gradients \citep[e.g.,][see also Fig.~\ref{fig:fig4}]{guendelman_atmospheric_2019}, which result in a significantly different circulation \citep{KANG2019}.

\begin{figure*}[htb!]
    \centering
    \includegraphics[width=16cm]{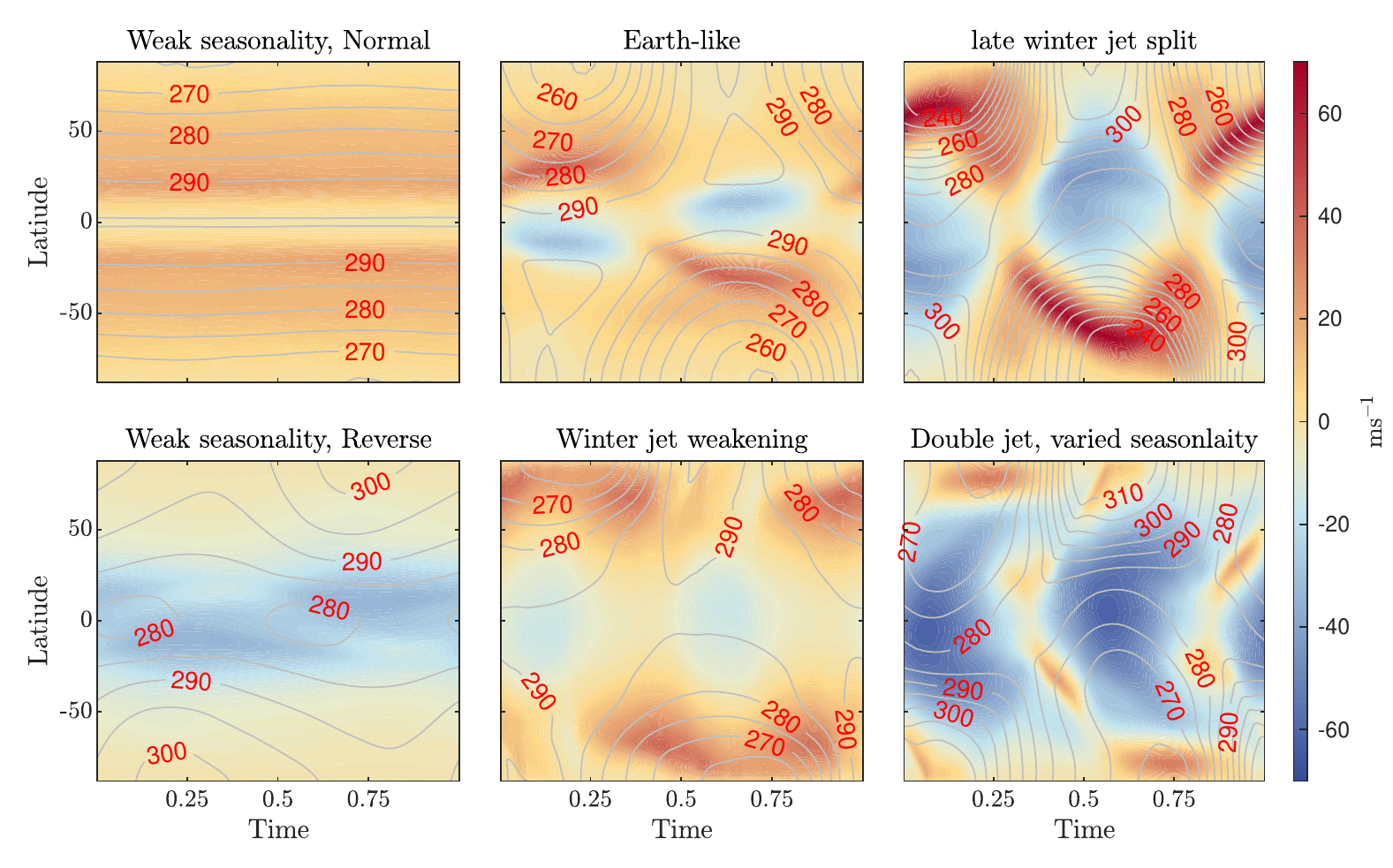}
    \caption{The zonal mean zonal wind vertically averaged between $\sigma=0.25$ and $\sigma=0.5$ (shading), and surface temperature (contours, contour interval $5$K) for the different jet seasonal cycle regimes detected in the parameter space. These panels are the ones highlighted in Figure~\ref{fig:fig2}. The details of each case are summarized in the text.}
    \label{fig:fig3}
\end{figure*}

In addition to variations in the strength of the winter vortex, the seasonal evolution of the polar vortex depends on the planetary parameters (Fig.~\ref{fig:fig2}). To show this, we focus our attention on several representative subspaces of our parameter space. First, consider the regime with short orbital periods (Fig.~\ref{fig:fig2}A). In this regime, the seasonal variability is weak and the annual mean climate dominates. At low obliquities, the polar vortex is in the midlatitudes (bottom panel in Fig.~\ref{fig:fig2}A); as obliquity increases, the vortex weakens to a point where it no longer exists, and instead of a westerly jet, there is an easterly jet. This is a result of the annual mean insolation gradient dependence on obliquity, where it weakens and reverses as obliquity increases \citep{guendelman_atmospheric_2019}.

Longer orbital periods result in an increase in the seasonal variability and the strength of the jet (Fig.~\ref{fig:fig2}B). The jet's seasonal cycle becomes more complex as the orbital period is increased, both in terms of its strength and latitudinal position. In cases with intermediate-to-fast rotation rates and long enough orbital periods (e.g. $\Omega \geq 1/2$, $\omega > 1/2$ in Fig.~\ref{fig:fig2}B), the jet occurs solely between fall and spring, and is strongest and in its most poleward position during midwinter. Additionally, when the orbital period becomes very long, the jet splits into two separate jets during late winter (top panel in Fig.~\ref{fig:fig2}B).

Consider now the jet variations with the rotation rate (Fig.~\ref{fig:fig2}C). In the Earth-like case, i.e., fast rotation rate and moderate obliquity (top panel in Fig.~\ref{fig:fig2}C), the dominant jet occurs during winter with only small variations in its position. When decreasing the rotation rate (going down in Fig.~\ref{fig:fig2}), the jet shifts poleward, and its strength varies non-monotonically. At slow rotation rates, the jet develops a unique seasonal cycle where the wind speed weakens during midwinter (bottom panel in Fig.~\ref{fig:fig2}C). A similar transition occurs for higher obliquity (Fig.~\ref{fig:fig2}D). At intermediate rotation rates with higher obliquities, the jet shows a complex seasonal cycle, where the wind speed is non-monotonic during winter, and there is a jet split during late winter (for example, second panel from the top in Fig.~\ref{fig:fig2}D). In addition, at high obliquities and slow rotation rates, the summer polar vortex is more dominant than its winter counterpart \citep[][bottom panel in Fig.~\ref{fig:fig2}D]{guendelman_2021}.

Figure~\ref{fig:fig3} shows six specific cases that highlight the different regimes of the jet seasonal cycle detected within the parameter space. The regimes are as follows:
\begin{enumerate}
    \item Weak seasonality, normal climate - The jet is in the midlatitudes with weak-to-no seasonality. This occurs when the obliquity is low or moderate and the orbital period is short (e.g., $\omega=1/8$, $\gamma=$ $10^{\circ}$ and $\Omega=1$). 
    \item Weak seasonality, reverse climate - No polar vortex, easterly winds in the low- and mid-latitudes with weak seasonality. This occurs when the obliquity is high and the orbital period is short (e.g., $\omega=1/8$, $\gamma=$ $90^{\circ}$ and $\Omega=1$). 
    \item Earth-like - The jet strengthens and shifts slightly poleward during midwinter. This occurs when the obliquity is moderate and the rotation rate is fast (e.g., $\omega=1$, $\gamma=$ $30^{\circ}$ and $\Omega=1$). 
    \item Winter jet weakening - The jet weakens during midwinter, accompanied by a slight poleward shift of the jet. This occurs when the obliquity is moderate or high and the rotation rate is slow (e.g., $\omega=1$, $\gamma=$ $30^{\circ}$ and $\Omega=1/16$). 
    \item Late winter jet split - The jet strengthens and shifts poleward during winter, and during late winter/spring, the jet splits into two jets. This occurs when the obliquity is moderate, the rotation rate is moderate or fast, and the orbital period is long (e.g., $\omega=4$, $\gamma=$ $30^{\circ}$ and $\Omega=1/2$). 
    \item Double jet with varied seasonality - There exists subtropical and polar jets, each having a different seasonality. This occurs when the obliquity is high and the rotation rate is moderate (e.g., $\omega=1$, $\gamma=$ $70^{\circ}$ and $\Omega=1/2$). 
\end{enumerate}

In the remainder of this section, we will examine more closely the dynamics of these different regimes.

\subsection{Weak seasonal cycle}

\begin{figure}[htb!]
    \centering
    \includegraphics[width=8cm]{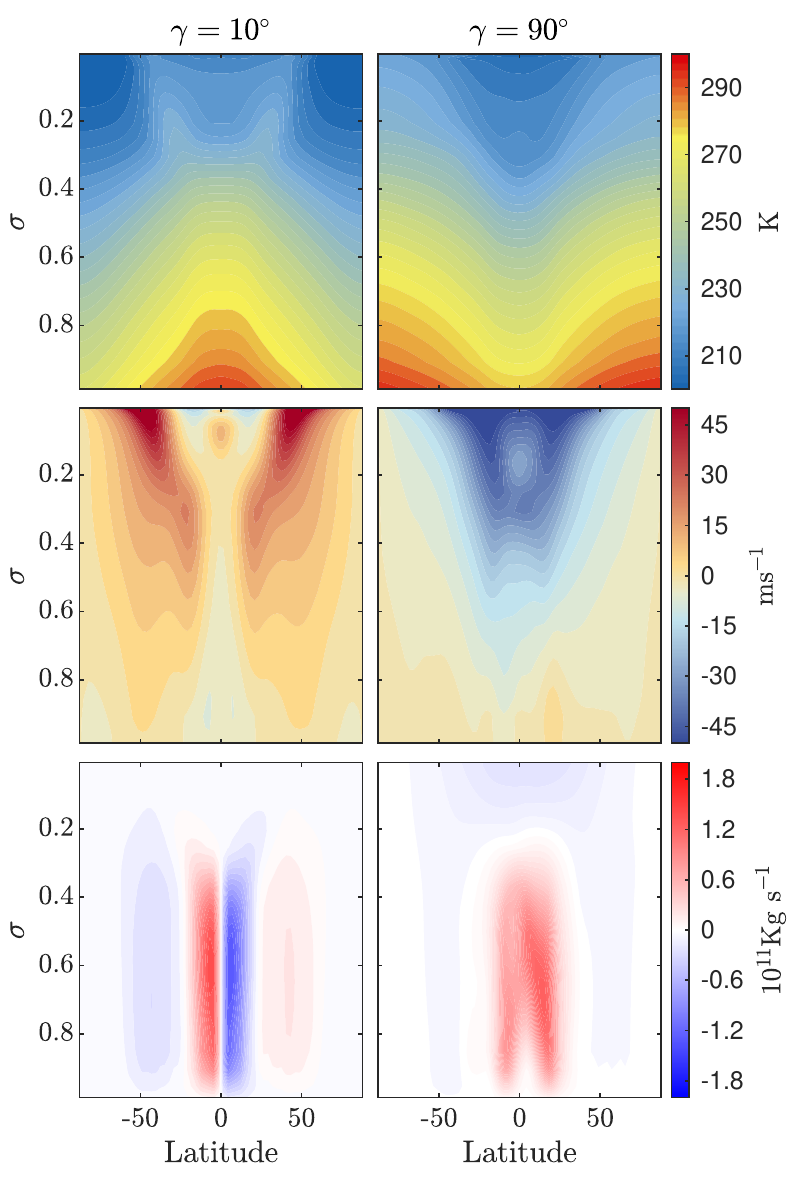}
    \caption{An equivalent June-August temporal mean of the vertical structure of temperature (K, first row), zonal mean zonal wind (ms$^{-1}$, second row), and mass streamfunction (Kg s$^{-1}$, third row) that represents the mean meridional circulation (red denotes counterclockwise circulation) for $\gamma = 10^{\circ}$ (left column) and $\gamma = 90^{\circ}$  (right column) with $\Omega=1$ and $\omega=1/8$.}
    \label{fig:fig4}
\end{figure}

First, we consider the regimes with a weak seasonal cycle, which in our simulations occur when the orbital period is short. The climate and atmospheric circulation on planets with a short enough orbital period, such that the annual mean forcing dominates, strongly depends on the planetary obliquity, which influences the latitudinal distribution of the annual mean insolation.

There are two processes that can accelerate and maintain a jet. The first is an equator-to-pole temperature differences that cause a meridional circulation from warm to cold latitudes. The mean meridional circulation is represented using the mass streamfunction, where the circulation follows the streamlines (blue represents a clockwise circulation and red a counterclockwise circulation); hereinafter, we will use the terms mean meridional circulation and streamfunction interchangeably. An air parcel that flows from low to high latitudes gains angular momentum, accelerating the jet. A jet that is maintained through this process is called a thermally-driven jet \citep{vallis_2017}. The second process that accelerates a jet occurs in baroclinic unstable regions. Waves that develop in these regions converge momentum towards the disturbance regions and accelerate an eddy-driven jet \citep{vallis_2017}. On Earth, these processes occur in proximity, and the resulting jet is called a merged jet; however, the jet changes its characteristics during the seasonal cycle \citep{lachmi2014, yuval_relation_2018}.

At low obliquities ($\gamma < 54^{\circ}$), the maximum annual mean insolation and temperature are at the equator, and the circulation is similar to Earth's annual mean circulation (Fig.~\ref{fig:fig4}). In each hemisphere there is a Hadley cell in the tropics, a Ferrel cell in the midlatitudes, and a jet between these two cells  \citep[e.g.,][]{vallis_2017}. This jet is a merged jet, i.e., a merger between the subtropical jet at the edge of the Hadley circulation and an eddy-driven jet in the Ferrel cell \citep[e.g.,][]{lachmi2014}. 

\begin{figure*}[htb!]
    \centering
    \includegraphics[width=16cm]{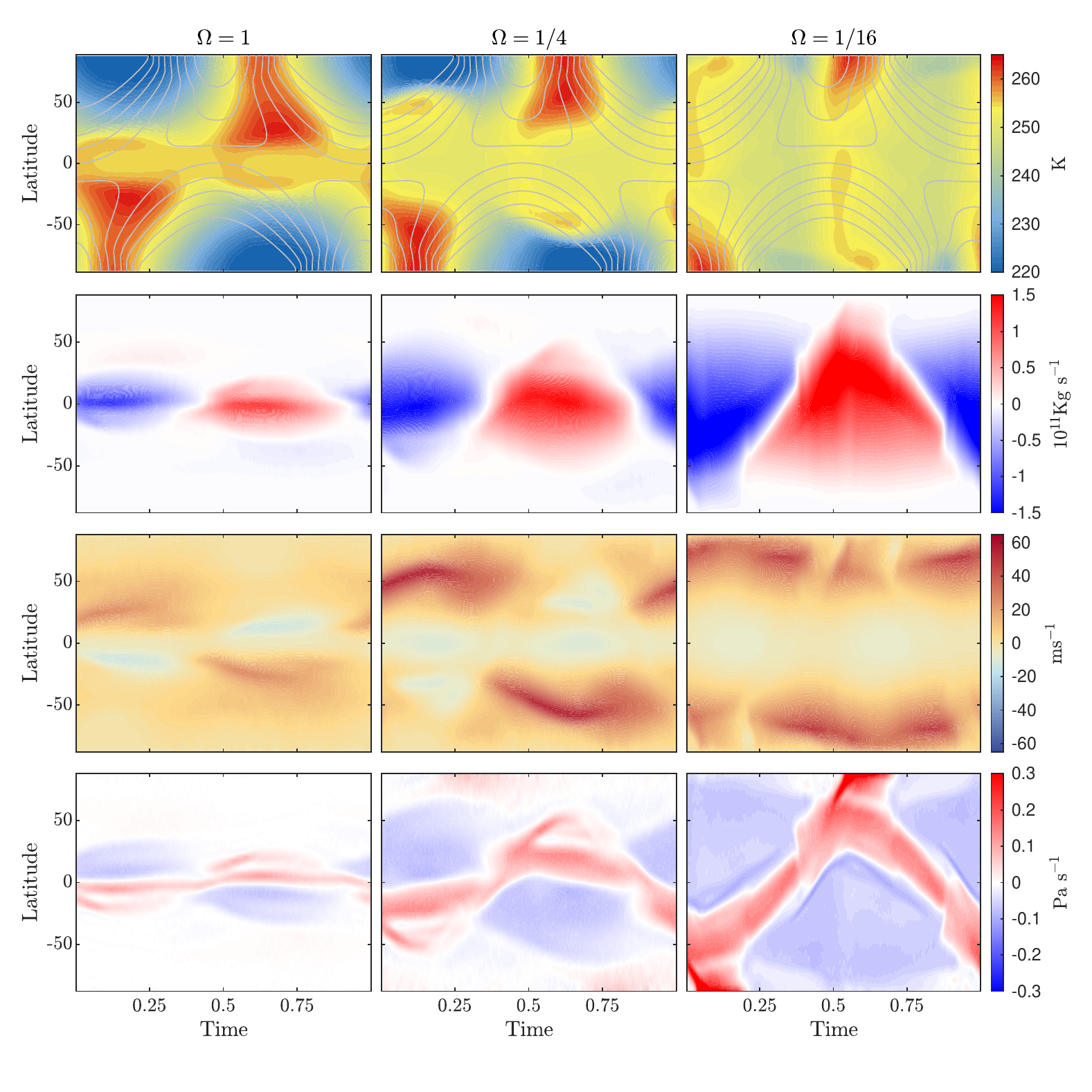}
    \caption{Comparison of the seasonal cycle of three simulations, $\Omega = 1, 1/4$ and $1/16$ with $\omega=1$, $\gamma=30^{\circ}$. Top row: Temperature at $\sigma=0.5$ (K, shading) and the pattern of the top of the atmosphere forcing (Wm$^{-2}$, contours). Second row: The mass streamfunction at $\sigma=0.5$ (Kg s$^{-1}$, red is southward flow). Third row: The zonal mean zonal wind at $\sigma=0.5$ (ms$^{-1}$). Fourth row: vertical velocity at $\sigma=0.5$ (Pa s$^{-1}$).}
    \label{fig:fig5}
\end{figure*}

In contrast, in high-obliquity ($> 54^{\circ}$) there is a reversal of the annual mean insolation gradients and the minimum temperature is at the equator. Instead of air ascending at the equator and descending at the midlatitudes, there is a cross-equatorial circulation. The warmer hemisphere dictates the direction of that cross-equatorial circulation, i.e., the cross-equatorial flow will be from the summer hemisphere (northern hemisphere in Fig.~\ref{fig:fig4}) to the winter hemisphere (southern hemisphere in Fig.~\ref{fig:fig4}). This suggests that, unlike the low-obliquity case, where a small deviation in the temperature field results in small deviations in the circulation, in the high-obliquity case, small changes in the temperature field result in significant variations in the circulation. Furthermore, the annual mean circulation in the high-obliquity case does not appear during the seasonal cycle, and is an artifact of the annual averaging. Unlike the meridional circulation, the zonal mean zonal wind has only small variations during the seasonal cycle. It consists of weak westerlies close to the surface, and no westerlies, i.e., no polar vortex, at high altitudes.

\subsection{Winter jet weakening}

For Earth-like conditions, the jet strengthens during the winter, reaching its maximum in midwinter. However, for planets with a slow enough rotation rate (i.e., $\Omega\leq0.25$) and a strong enough seasonal cycle (high obliquity and long orbital period, Figs.~\ref{fig:fig2}C-D), the winter jet weakens during midwinter. As the seasonal cycle becomes stronger, the summer jet weakening occurs at a faster rotation rate (Figs.~\ref{fig:fig2}C-D).

\begin{figure*}[htb!]
    \centering
    \includegraphics[width=16cm]{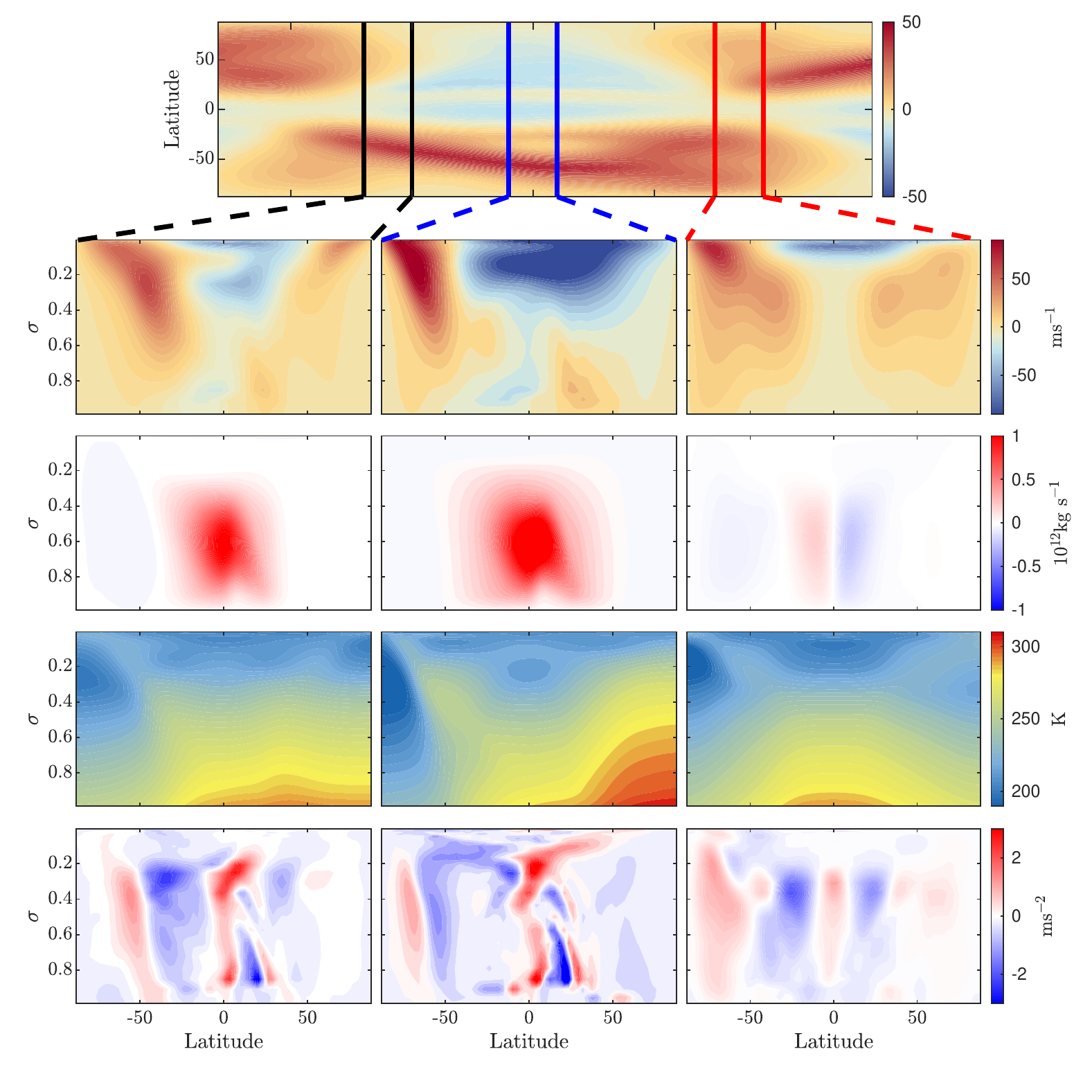}
    \caption{Top row: Seasonal variations of the zonal mean zonal wind (at a height of $\sigma=0.5$) around southern hemisphere winter for $\omega=4$, $\gamma=30^{\circ}$ and $\Omega=0.5$. As in Fig.~\ref{fig:fig2}, the x-axis represents fraction of year. The fractions confined between the colored vertical lines represent the time average period for the respective columns below it: black (left column) for early winter (late fall), blue for midwinter and red for late winter (early spring). Second row: Vertical structure of the zonal mean zonal wind (ms$^{-1}$). Third row: Vertical structure of the mass streamfunction ($10^{12}$ Kg s$^{-1}$). Fourth row: Vertical structure of the temperature (K). Fifth row: Vertical structure of the eddy momentum flux convergence (ms$^{-2}$).}
    \label{fig:fig6}
\end{figure*}

To further understand the seasonal transition to a midwinter weakening of the jet, we compare the seasonal evolution of three cases with different rotation rates ($\Omega=1,/ 1/4, / 1/16$) and otherwise identical parameters ($\omega=1$ and $\gamma=30^{\circ}$). Comparing the latitudinal structure of the temperature at $\sigma=0.5$ and streamfunction (top two rows in Fig.~\ref{fig:fig5}), shows that regions with a weak meridional temperature gradient correspond to regions with a strong streamfunction. The weak temperature gradients are the result of efficient heat transport by the mean meridional circulation. In addition, the high-level temperature in the ascending and descending regions of the Hadley circulation are warmer than other latitudes (top row in Fig.~\ref{fig:fig5}). The reason for this warming differs for the ascending and descending regions. The ascending region is warmer because of the stronger radiative and latent heating that occur in the summer hemisphere, while the descending region is warmer because of the combined effect of the Hadley cell heat transport and adiabatic heating due to the descending motion in this region. 

The main difference between the regime where the jet strengthens during midwinter (fast or intermediate rotation rates, two left columns in Fig.~\ref{fig:fig5}) and the regime with a midwinter weakening of the jet (slow rotation rates, right column in Fig.~\ref{fig:fig5}) is that the Hadley cell extends from one pole to the other in the latter case, but not in the former one (third row in Fig.~\ref{fig:fig5}). In cases with fast and intermediate rotation rates, where the descending edge does not reach the winter pole, there is a strong temperature gradient at high latitudes due to radiative cooling around the pole, as these regions receive low-to-no radiation during winter (see insolation patterns in the top row of Fig.~\ref{fig:fig5}). These meridional temperature gradients maintain the strong jet during midwinter. However, this is not the case at slow rotation rates, where the temperature gradient is weak as a result of the expansion of the Hadley circulation and the descending motion being close to the winter pole (right column in Fig.~\ref{fig:fig5}). 

Examining the seasonal evolution shows that in cases with fast rotation rates, where the Hadley circulation width is constrained \citep{guendelman_axisymmetric_2018, hill2019, singh2019, lobo_atmospheric_2020}, strong cooling occurs in the winter hemisphere. The edge of the Hadley cell, which can be inferred from the temperature field at $\sigma=0.5$ as a warm patch in the winter hemisphere, is in the jet's vicinity. For an Earth-like rotation rate, where the Hadley cell covers the tropics, the meridional temperature gradients are not as sharp as the temperature gradients in the $\Omega=1/4$ case, where the descending edge is more poleward. During the seasonal cycle, these gradients become even sharper, and can maintain a stronger jet (Fig.~\ref{fig:fig5}). However, in the slow rotation rate case, the cooling of polar latitudes occurs only in short periods during spring and autumn, where descending motion does not reach the poles (right panel in the bottom row in Fig~\ref{fig:fig5}). In contrast, during winter, the descending motion is in the polar regions, diminishing the meridional temperature gradients, which results in a weaker jet (right panel in the bottom row in Fig.~\ref{fig:fig5} around $t=0.5$).

To summarize, for slow rotation rates and a strong enough seasonal cycle, the Hadley circulation extends from one pole to the other during midwinter. The wide circulation and the descending motion around the pole, results in weak meridional temperature gradients and a weak polar vortex during midwinter. This is not the case when it comes to the transition seasons or faster rotation rates. In these cases, the descending motion is equatorward of the pole, and the primary process that occurs at the pole is radiative cooling. The cooling in the polar regions creates a strong temperature gradient that maintains a strong polar vortex (Fig.~\ref{fig:fig5}). 

\subsection{Late winter jet split}

Next, we consider the regime where the jet splits during the transition seasons. This occurs for long orbital periods, intermediate obliquity, and intermediate-to-fast rotation rates. Figure~\ref{fig:fig6} shows the detailed evolution of the zonal mean zonal wind, mean meridional circulation, temperature and eddy momentum fluxes for this case. Although the most noticeable split occurs during the late winter-to-spring transition, there is a weaker jet split during the fall-to-early winter (e.g., see southern hemisphere in the rightmost column in Fig.~\ref{fig:fig6}). Both split jets result from similar dynamics that are related to the seasonal cycle of the Hadley circulation. In the transition seasons, the circulation is close to hemispherically symmetric, and due to the relatively fast rotation rate, the cells span only up to midlatitudes ($\sim 30^{\circ}$, rightmost column in Fig.~\ref{fig:fig6}), and the thermally driven jet is in the descending edge of the Hadley circulation. The polar latitudes during early and late winter are colder than the midlatitudes, resulting in steep meridional temperature gradients that maintain the eddy-driven jet there (Fig.~\ref{fig:fig6}). In contrast, during midwinter, the Hadley circulation becomes cross-equatorial, and its descending edge reaches to higher latitudes (middle column in Fig.~\ref{fig:fig6}). This results in a poleward shift of the thermally driven jet and merging with the eddy driven jet (Fig.~\ref{fig:fig6}). 
The split of the jet into its thermally- and eddy-driven components was observed in both Earth \citep{lachmi2014, yuval_relation_2018} and previous modelling studies. More specifically, a split jet was found in a previous modeling study of planets with a seasonal cycle due to non-zero eccentricity; in that study, the split of the jet was also related to long orbital periods \citep{guendelman_atmospheric_2020}. This possibly points to a characteristic eddy timescale that allows this phenomenon to develop mainly in long enough orbital periods. %However, this can also result from the model simplicity, and understanding the detailed dynamics and dependence is left for a future study.

\subsection{Double jet with varying seasonality}

\begin{figure*}[htb!]
    \centering
    \includegraphics[width=16cm]{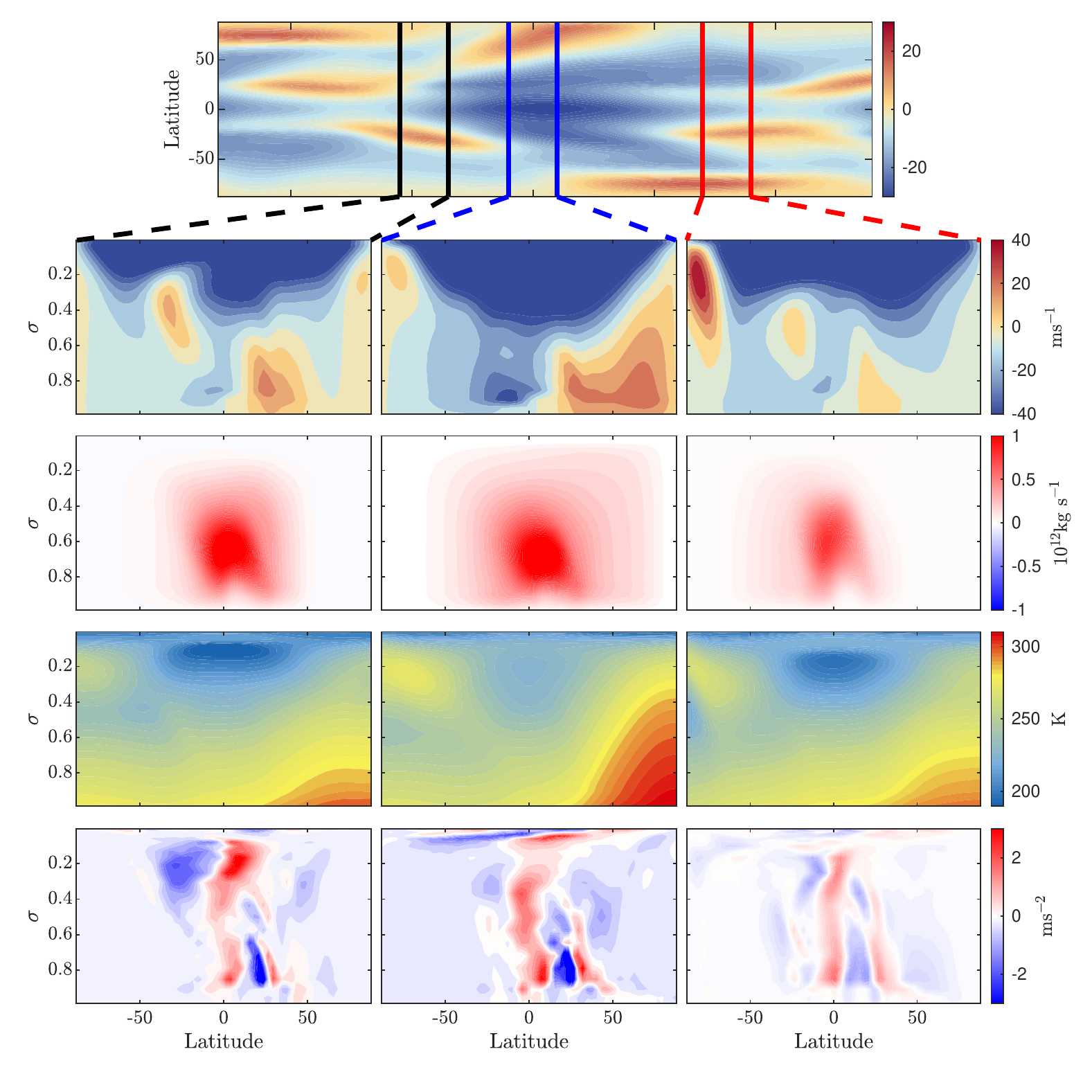}
    \caption{Top row: The zonal mean zonal wind (ms$^{-1}$) seasonal cycle around southern hemisphere winter for $\omega=1$, $\gamma=70^{\circ}$ and $\Omega=0.5$. The fractions confined between the vertical lines represent the time average period for the respective column below it: black for early winter (late fall), blue for midwinter, and red for late winter (early spring). Second row: The zonal mean zonal wind (ms$^{-1}$). Third row: mass streamfunction (Kg s$^{-1}$). Fourth row: The vertical structure of temperature (K). Fifth row: Eddy momentum flux convergence (ms$^{-2}$).}
    \label{fig:fig7}
\end{figure*}

At high obliquities and intermediate-to-slow rotation rates, the seasonality of the westerly winter jets are very complex (Fig.~\ref{fig:fig2}D). For the case shown in Fig.~\ref{fig:fig7}, the jet weakens during midwinter and splits during the transitions seasons. At first glance, this is simply a combination of the previous two regimes discussed above, the winter jet weakening and the late winter jet splitting. However, a deeper examination shows substantial differences in the vertical temperature structure and dynamics. As opposed to the low obliquity case, where in the upper atmosphere the polar latitudes are colder than the midlatitudes, in the high obliquity case, they are warmer. This leads to both an inversion in the vertical temperature profile (see the vertical temperature profile in the south polar regions in Fig.~\ref{fig:fig7} second row from the bottom) and a reversed meridional temperature gradient at the top of the atmosphere (Fig.~\ref{fig:fig7} second row from the bottom around the height of $\sigma=0.2$). This temperature structure and the fast radiative transitions that occur at high obliquities result in the complex seasonal cycle seen at high obliquities and intermediate rotation rates. 

During the autumn-to-early winter, there are two different jets in the winter hemisphere, one jet at the descending edge of the Hadley circulation (around latitude $\sim40$S) and a weaker high-level polar jet that results from the temperature gradients in the polar latitudes. During midwinter, the Hadley circulation widens, and the jets merge. Similar to the midwinter jet weakening regime discussed previously, the widening of the Hadley circulation results in weaker temperature gradients and a weaker winter jet (Fig.~\ref{fig:fig7}). In late winter and the beginning of spring, the Hadley circulation narrows, a jet forms in the descending region of the winter cell and separates from the polar jet that strengthens due to the cooling of the mid-troposphere (right column in Fig.~\ref{fig:fig7}). As a result of the complexity of the seasonal cycle and the response of the meridional circulation, the subtropical jet is centered in the mid-troposphere inside the Hadley cell (Fig.~\ref{fig:fig7}, around latitude $30^{\circ}$ and $\sigma=0.5$), rather than at the edge of the circulation. In addition, there is only a weak eddy momentum flux convergence in all the jets, suggesting that the jets are mainly thermally driven. 

In addition to the winter jets, in this case, we can see the dominance of the summer jet (middle panel in Fig.~\ref{fig:fig7}). This summer jet is discussed in detail in \cite{guendelman_2021}, and is related to the widening of the Hadley circulation and the increased efficiency of the vertical momentum transport as the rotation rate is decreased at high obliquity.

Finally, we note that in this regime, the complex seasonal cycle at high obliquities results from the temperature response to the radiative forcing. This could be the result of the simplified radiation scheme used in this model. However, \citet{Kang_2019b}, using a different model, had a temperature structure with similarities to the one showen here, with warmer poles at higher levels for high obliquity (see Fig.~4f there). This indicates that this is not model dependent but rather a robust response to the high obliquity radiative forcing. %These results need to be studied further to understand the role of radiation schemes and other parameters on the resulting climate.

\section{Relation between the jet and storm activity}\label{sec:EKE}

\begin{figure}[htb!]
    \centering
    \includegraphics[width=7cm]{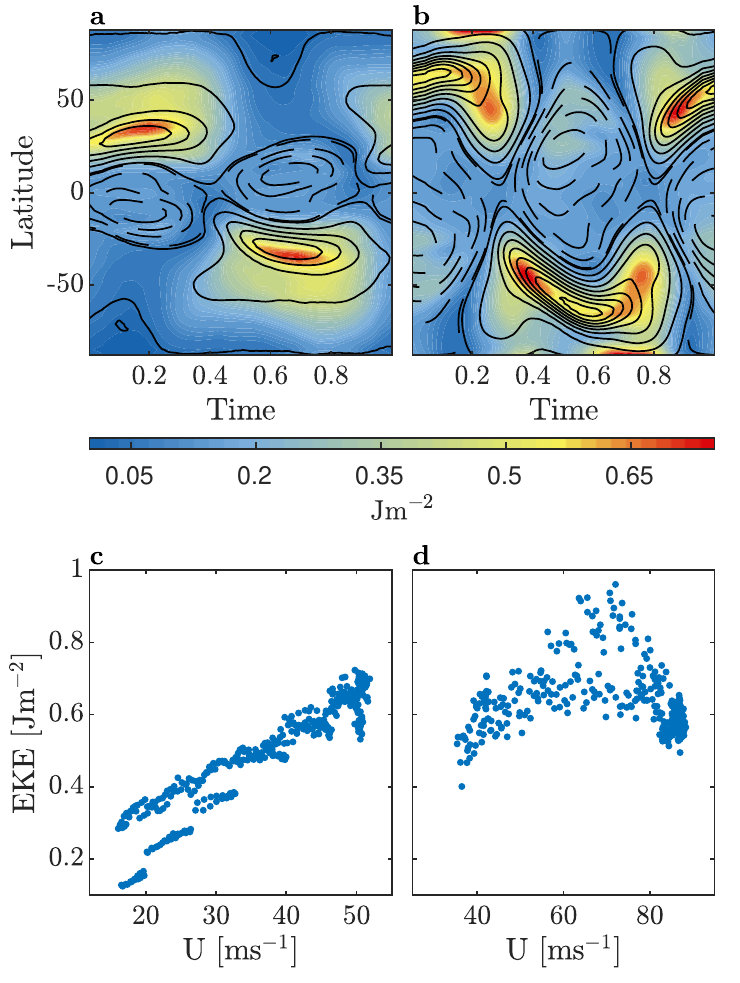}
    \caption{Top: Vertically integrated EKE (shading, $10^6\rm{Jm^{-2}}$, Eq.~\ref{eq:EKE}) and zonal mean zonal wind (contours, contour intervals are $10\rm{ ms^{-1}}$) seasonal cycle at a height of $\sigma=0.3$ for an Earth-like simulation (a) and for a simulation with $\omega=4$, $\gamma=30^{\circ}$ and $\Omega =1/2$ (b, the same simulation used for the late winter jet split). Bottom, scatter plot of vertically integrated EKE ($10^6\rm{Jm^{-2}}$) as a function of the jet strength ($\rm{ ms^{-1}}$) for an Earth-like simulation (c) and for a simulation with $\omega=4$, $\gamma=30^{\circ}$ and $\Omega =1/2$ (d). Each point represents the value of the vertically integrated EKE at the latitude of maximum zonal mean zonal wind in each day.}
    \label{fig:fig8}
\end{figure}

The jet regions usually collocate with regions of increased storm activity, a result of the increased baroclinicity in the jet region \citep{vallis_2017}. The storm activity is measured here as the variance of the total kinetic energy, expressed as the vertically integrated eddy kinetic energy (EKE)
\begin{equation}\label{eq:EKE}
   [\rm{EKE}] = \int\frac{u'^2+v'^2}{2}\frac{dp}{g},
\end{equation}
where $u'$ and $v'$ are the deviations from the zonal mean ($\overline{u},\overline{v}$) and the square brackets denote vertical integration. Linear baroclinic instability theory predicts that EKE will increase with increasing jet strength \citep{charney1947, Eady1949}. However, this is not the case in all our simulations. In some of our simulations, the EKE is, in fact, weaker during periods when the jet is strongest. This phenomenon is dispersed throughout the parameter space, occurring mainly in intermediate obliquity values with either long orbital periods or intermediate rotation rates, where the jet speed reaches high velocities. For simplicity, we use the example simulation of the late-winter jet split regime ($\Omega=1/2$, $\gamma=30^{\circ}$ and $\omega=4$) also as an example of an EKE-minimum simulation. However, the suppression of storm activity and the late-winter jet split do not always occur together. 

In both the Earth-like and EKE-minimum simulations, the jet reaches its maximum strength during midwinter (Figs.~\ref{fig:fig3} and \ref{fig:fig8}a-b). However, the two simulations differ in the response of the EKE. In the Earth-like simulation, the EKE follows the strengthening of the jet. In the EKE-minimum simulation, the EKE maximizes during the transition seasons and is weaker during midwinter. Focusing on the EKE-minimum scenario, during autumn-early winter, the jet is at the descending edge of the Hadley cell and can be characterized as a merged jet, where the poleward flank of the jet is eddy driven, which is evident from both the Ferrel cell close to the winter pole and the eddy momentum flux convergence there (Fig.~\ref{fig:fig6}). During midwinter, with the widening of the Hadley cell, the jet shifts poleward and strengthens. The eddy momentum flux convergence is weaker compared to the transition seasons (Fig.~\ref{fig:fig6}), as well as the EKE (Fig.~\ref{fig:fig8}b). During late winter, the jet shifts equatorward and splits into a merged jet and an eddy-driven jet (Fig.~\ref{fig:fig6} and \ref{fig:fig8}), and EKE starts to strengthen again.

A similar minimum of EKE occurs on Earth over the Pacific \citep{nakamura_midwinter_1992}; however, there are different characteristics between the two. First, on Earth, the jet shifts equatorward during midwinter, unlike our simulation, where it shifts poleward. In addition, the jet on Earth transitions from a merged jet during the transition season to a more subtropical, thermally driven jet during midwinter \citep[e.g.,][]{yuval_relation_2018}. Despite these differences in characteristics, the explanation for the EKE minimum on Earth proposed by \citet{hadas_2020}, who connected the decrease in EKE with a reduction in the number of storms and their lifetime due to a disconnect between the upper and lower levels as the jet strength increases can also apply to our simulations, given the similar jet speed dependence. When comparing the Earth-like and EKE-minimum simulations, at lower jet speeds we can see a similar jet strength dependence (Fig.~\ref{fig:fig8}c-d). However, the jet in the EKE-minimum simulation reaches higher speeds than in the Earth-like simulation, and surpasses a threshold after which EKE starts to decrease with increasing jet strength (Fig.~\ref{fig:fig8}c-d). This holds more generally, with a midwinter minimum in EKE occurring in simulations where the jet speed exceeds around 60-70 ms$^{-1}$. In addition, as the jet strengthens, it also narrows (Fig.~\ref{fig:fig8}b), and \citet{harnik_effects_2004} have shown that as the jet narrows, there is a decrease in the meridional wavelength of the perturbation resulting in the perturbations growing less, and this can also explain the EKE-minimum.

\section{Potential vorticity structure}\label{sec:PV}

\begin{figure}[htb!]
    \centering
    \includegraphics[width=7.5cm]{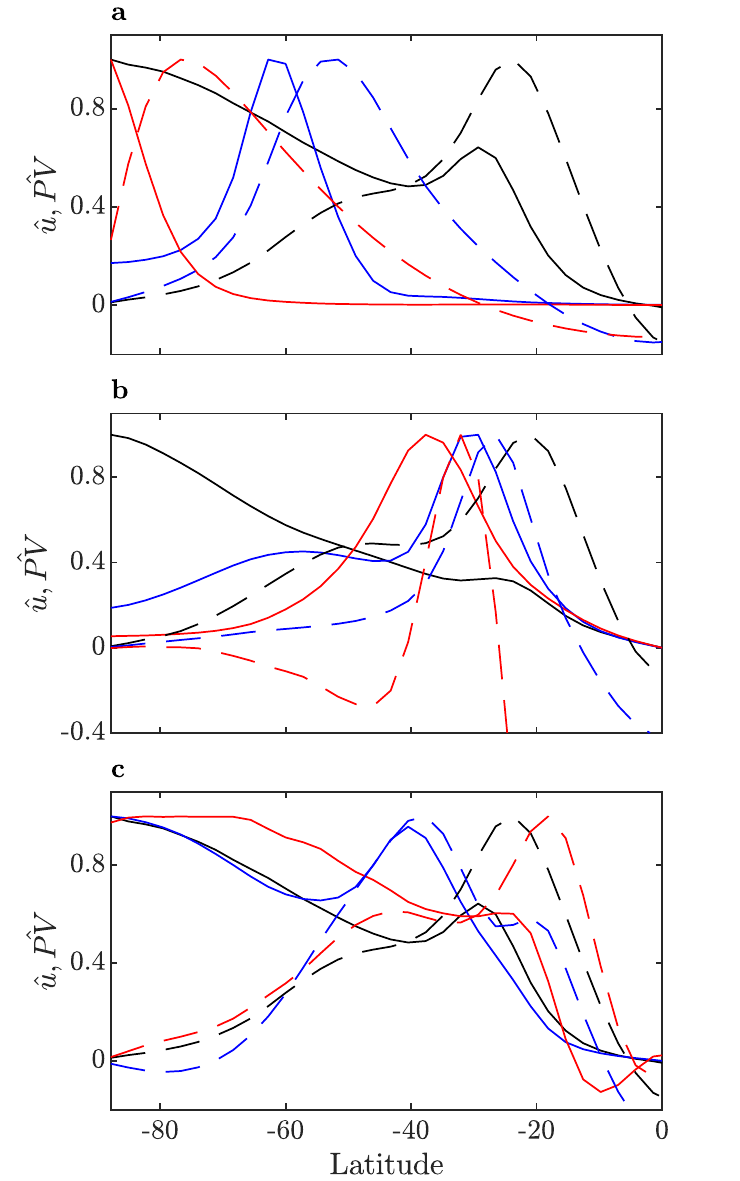}
    \caption{Absolute value PV (solid) and zonal mean zonal wind (dashed) meridional structure in the winter hemisphere, for different parameter values, for a temporal mean of days 180-240 and $\sigma=0.5$. Each line is normalized by its maximum value. (a) black is for $\Omega=1 $, blue is for $\Omega=1/4$ and red is for $\Omega=1/16$; other parameters are constant, with $\omega=1$ and $\gamma=30^{\circ}$. (b) black is for $\gamma=20^{\circ}$, blue is for $\gamma=50^{\circ}$ and red is for $\gamma=80^{\circ}$; other parameters are constant, with $\omega=1$ and $\Omega=1$. (c) black is for $\omega=1 $, blue is for $\omega=4$ and red is for $\omega=1/4$; other parameters are constant, with $\Omega=1$ and $\gamma=30^{\circ}$.}
    \label{fig:fig9}
\end{figure}

The potential vorticity (PV) is a useful quantity when studying the dynamics of vortices \citep{hoskins1985}, as in the absence of diabatic forcing and friction, PV acts as a conserved tracer. We calculate the PV using 
\begin{equation}\label{eq:PV}
    \rm{PV} = -g(f+\zeta)\frac{\partial\theta}{\partial p},
\end{equation}
where $g$ is the surface gravity, $f=2\Omega\sin\phi$, $\phi$ is latitude, $\theta$ is the potential temperature, and $\zeta$ is the vertical component of the relative vorticity given by
\begin{equation}
    \zeta = \frac{1}{a\cos\phi}\frac{\partial v}{\partial\lambda} - \frac{1}{a\cos\phi}\frac{\partial}{\partial\phi}(u\cos\phi),
\end{equation}
where $\lambda$ is longitude. Because PV can vary significantly in its vertical structure, we follow previous studies \citep[e.g.,][]{lait94, mitchell_polar_2015, waugh_martian_2016, sharkey_potential_2021} and use the scaled PV given by 
\begin{equation}
    \rm{PV}_s = \rm{PV}\left(\frac{\theta}{\theta_0}\right)^{-(1+R/c_p)},
\end{equation}
where $\theta_0$ is a reference potential temperature. This form removes a significant amount of the PV vertical variation.

In Earth's atmosphere, PV generally monotonically increases (in absolute values) from the equator to the poles. However, this is not the case on Mars \citep[e.g.,][]{mitchell_polar_2015} or Titan \citep{sharkey_mapping_2020,sharkey_potential_2021}, where the PV has an annular structure, i.e., it maximizes equatorward from the poles. The persistence of an annular PV structure is somewhat surprising, as past studies have shown that an annular PV structure can be barotropically unstable \citep{dritschel_roll-up_1992}. Given this, studies of the Martian polar vortex have suggested that forcing is needed to maintain the annular structure \citep[e.g.,][]{mitchell_polar_2015}. \citet{toigo_what_2017} suggested that latent heating from CO$_2$ condensation is what forces the annular PV structure on Mars. In addition to that, \citet{scott_forcing_2020} have shown, using a simplified shallow water model, that annular PV, can be stable and be maintained solely by transport from the Hadley circulation. 

To examine the dependence of the PV structure on the different planetary parameters, it is insightful to compare the latitudinal structures of PV together with that of the zonal mean zonal wind. For an Earth-like configuration (black lines in Fig.~\ref{fig:fig9}a), the jet is at around $20^{\circ}$S and the strongest PV meridional gradients occur close to this latitude. There is a local maximum in the PV near the jet, but it is not a global maximum, and the PV maximizes at the pole (black lines in Fig.~\ref{fig:fig9}a). Decreasing the rotation rate results in two main responses: the jet shifts poleward, and the background vorticity (i.e., the Coriolis parameter $f$) weakens. The combination of these changes results in a global maximum of PV poleward of the jet, with the maximum PV gradient is still close to the maximum winds. In intermediate rotation rates, where the jet is not too close to the pole, this results in an annular PV (e.g., $\Omega=1/4$, blue curves in Fig.~\ref{fig:fig9}a). For slow rotation rates, the jet is close to the pole and the maximum PV is again at the pole (e.g., $\Omega=1/16$, red curves in Fig.~\ref{fig:fig9}a).

Increasing the obliquity shifts the jet poleward and makes it narrower (dashed curves in Fig.~\ref{fig:fig9}b). This, in turn, increases the effect of the relative vorticity, allowing, in some cases, an annular PV to persist (Fig.~\ref{fig:fig9}b). Increasing the orbital period also results in a shift poleward of the vortex; however, the narrowing effect seen as the orbital period increases is less dominant (Fig.~\ref{fig:fig9}c).

There is a trade-off between increasing obliquity and decreasing rotation rate when considering the polar vortex PV structure. For fast rotation rates, cases with high obliquity have annular PV (Fig.~\ref{fig:fig9}b), and as the rotation rate decreases, it shifts to a monopolar PV, as the polar vortex is closer to the pole (Fig.~\ref{fig:fig9}a). An opposite effect occurs at low obliquity values. For fast rotation rates, the PV is monopolar, while for the same obliquity value, an annular PV can persist in slower rotation rates, up to the point where the vortex reaches close to the pole (Fig.~\ref{fig:fig11})

\begin{figure}[htb!]
    \centering
    \includegraphics[width=7.5cm]{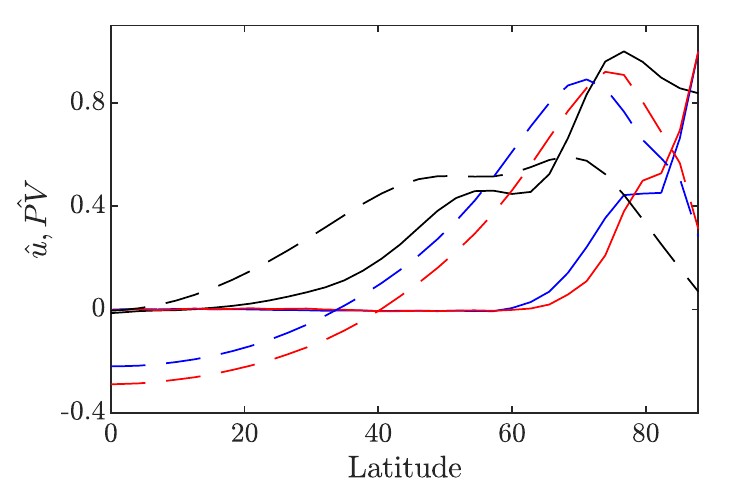}
    \caption{PV (solid) and zonal mean zonal wind (dashed) meridional structure in the summer hemisphere for different parameter values for a temporal mean of days 180-240, vertically averaged between $\sigma=0.5$ and $\sigma=0.85$. Each line is normalized by its maximum value in the summer hemisphere. Black is for $\gamma=20^{\circ}$, blue is for $\gamma=50^{\circ}$ and red is for $\gamma=80^{\circ}$; other parameters are constant, with $\omega=1$ and $\Omega=1/16$.}
    \label{fig:fig10}
\end{figure}

The above analysis focused on the winter polar vortex. However, for slow rotation rates and high obliquity, the dominant jet is in the summer hemisphere \citep{guendelman_2021}. Because this jet is centered in the low-mid troposphere, we examine the PV structure of the summer hemisphere polar vortex at lower vertical levels for different values of obliquity (Fig.~\ref{fig:fig10}). For low obliquity, the low-level jet is not fully developed, and the dominant jet in the summer hemisphere is the subtropical jet. Due to the slow rotation, the PV has an annular structure (black lines in Fig.~\ref{fig:fig10}), similar to the winter polar vortex in slow rotation rates and low obliquity. For higher obliquities (blue and red lines in Fig.~\ref{fig:fig10}), the low-level summer jet dominates, and there is a step-like structure, where PV in the poleward flank of the vortex is constant and there are steep PV gradients equatorward and poleward of it.

\begin{figure}[htb!]
    \centering
    \includegraphics[width=7.5cm]{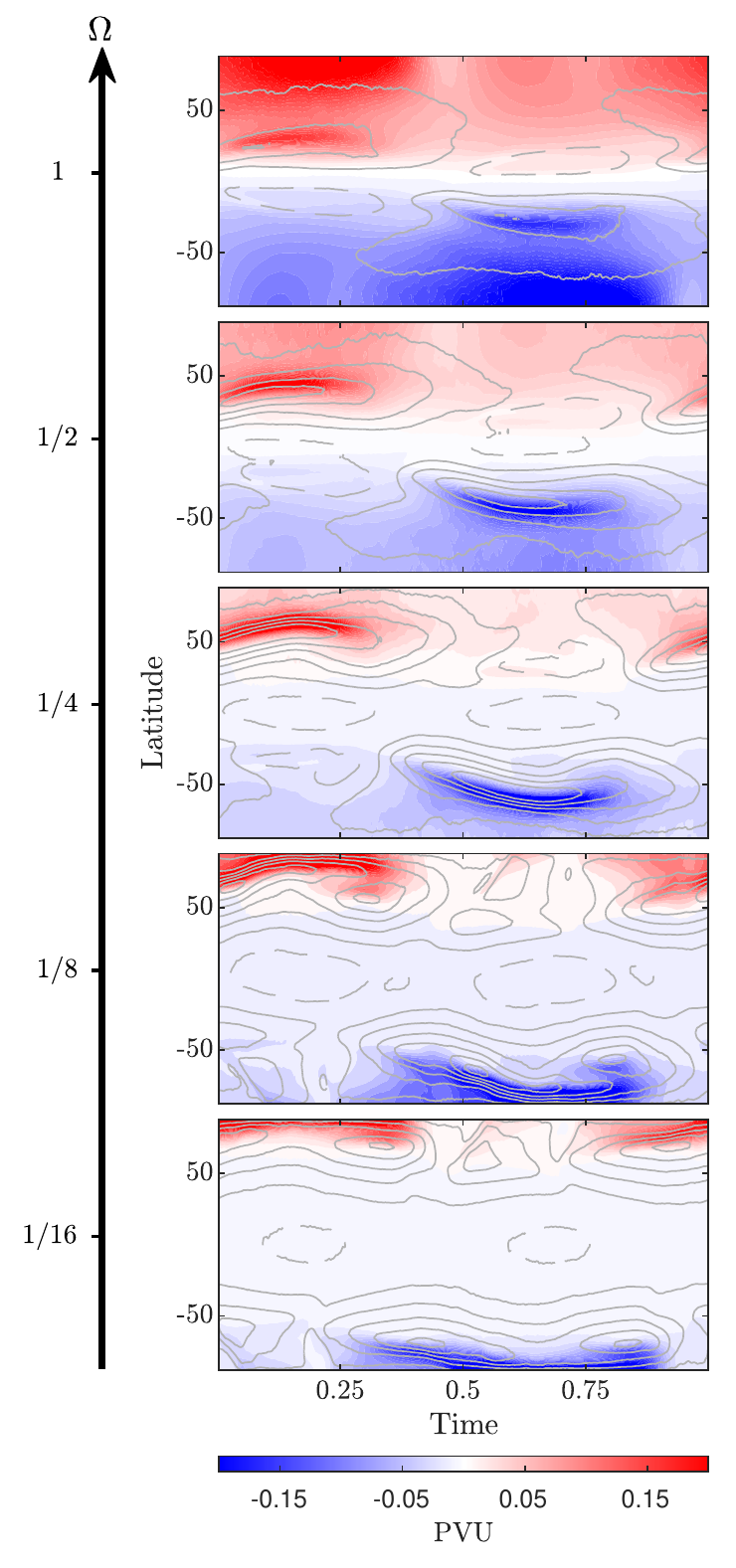}
    \caption{The potential vorticity (shading, PVU, i.e., $10^{-6}$ $\rm{K m^2 s^{-1} Kg^{-1}}$) and zonal mean zonal wind (contours, contour interval $10 \rm{m s^{-1}}$) seasonal cycle at a height of $\sigma = 0.5$ for different values of rotation rates ($\Omega$) with $\gamma=30^{\circ}$ and $\omega =1$.}
    \label{fig:fig11}
\end{figure}

The PV characteristics changes not only with the planetary parameters, but also during the seasonal cycle, and the PV can shift between monopolar and annular during the year. Fig~\ref{fig:fig11} shows the seasonal cycle of PV (shading) and zonal mean zonal wind for different values of rotation rate with $\gamma=30^{\circ}$ and $\omega=1$. At a fast rotation rate ($\Omega=1$, top panel in Fig.~\ref{fig:fig11}), similar to the seasonal mean, there is a local PV maximum in the jet's vicinity, however the global maximum is at the pole. At lower rotation rates, there are cases where the polar vortex has an annular PV during the entire winter ($\Omega=1/2$ second panel in Fig.~\ref{fig:fig11} ). However, for slow rotation rates ($\Omega\leq 1/4$, bottom three panels in Fig.~\ref{fig:fig11}) the polar vortex PV shifts between annular and monopolar, similar to what is seen in modelling studies of Titan \citep{shultis_2022}.

\begin{figure}[htb!]
    \centering
    \includegraphics[width=7.5cm]{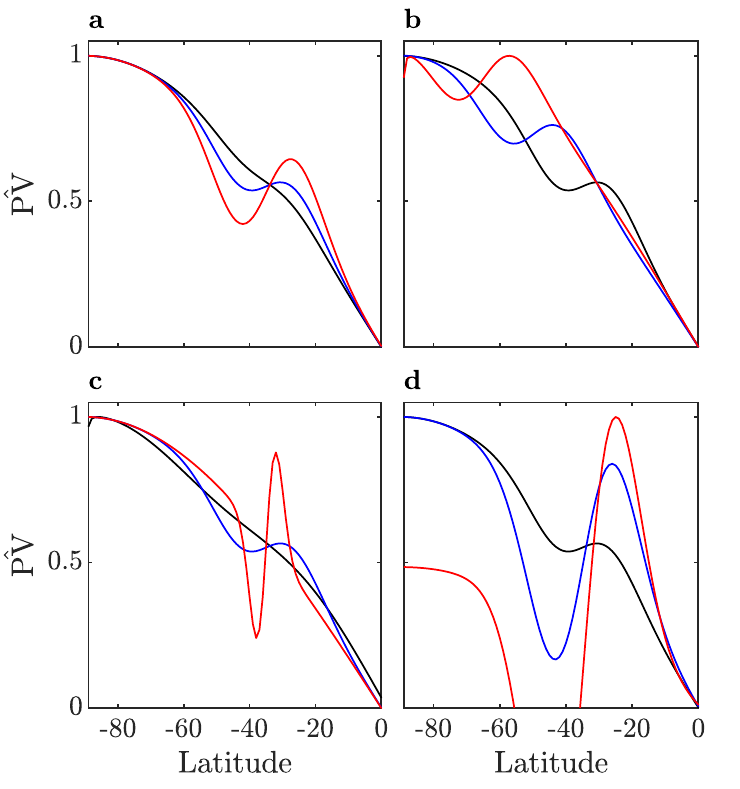}
    \caption{The value of $f+\overline{\zeta}$ normalized (each line is normalized by its maximum) using Eq.~\ref{eq:gauss} for the zonal mean zonal wind for different vortex strengths (a, $U_0=10$, black, $U_0=30$, blue, $U_0=60$, red), latitude (b, $\phi_0=35^{\circ}$, black, $\phi_0=50^{\circ}$, blue, $\phi_0=65^{\circ}$, red), width (c, $\phi_w=20^{\circ}$, black, $\phi_w=10^{\circ}$, blue, $\phi_w=3^{\circ}$, red) and rotation rates (d, $\Omega=1$, black, $\Omega=1/4$, blue, $\Omega=1/16$, red). Default parameter values: $U_0=30$, $\phi_0=35^{\circ}$, $\phi_w=10^{\circ}$, $\Omega=1$.}
    \label{fig:fig12}
\end{figure}

To get a better intuition for why there exists an annular PV in some parts of the parameter space, it is insightful to consider the case where $\partial_p\theta$ is approximately constant. The main variations in the zonal-mean PV are then due to the competing effects of the planetary vorticity, i.e., the Coriolis term $f=2\Omega\sin\phi$ and the zonal-mean relative vorticity $\overline{\zeta}= \frac{1}{a\cos\phi}\frac{\partial}{\partial\phi}(\overline{u}\cos\phi)$. The rotation rate is the only parameter that contributes to changes in the background planetary vorticity, but all the parameters can cause changes in the meridional structure of the zonal wind (and hence the relative vorticity). Changes in the zonal wind can be expressed as changes in the latitude, strength, and width of the jet. To better understand the role of these jet characteristics on the PV structure, we assume a Gaussian jet of the form
\begin{equation}\label{eq:gauss}
    U=U_0\exp\left[-\frac{(\phi+\phi_0)^2}{2\phi_w^2}\right].
\end{equation}
This corresponds to a jet centered around latitude $-\phi_0$ with a characteristic width of $\phi_w$ and a strength of $U_0$. The effect of independent changes in the jet's strength, latitude, and width, as well as rotation rate are shown in Fig.~\ref{fig:fig12}. The rotation rate has the most significant effect on the PV structure, with an annular PV occurring at a low rotation rate. The changes in jet characteristics have a comparable impact on the meridional structure of the PV near the jet, but whether this causes an annular structure (higher or lower values at the pole than at the jet) depends on the characteristics. Changing the strength or width of the jet alters the PV gradients at the jet, and can lead to a local maximum, but not to an annular vortex (for jets centered around $35^{\circ}$, Fig.~\ref{fig:fig12}a,c). However, when the jet is moved towards the pole, it can form an annular PV structure, especially when the jet is at high latitudes, but not too close to the pole (Fig.~\ref{fig:fig12}b). This is a result of the weak planetary vorticity gradients at high latitudes, which means a relatively small effect from $\zeta$ can result in an annular structure. Note that this analysis is highly simplified, but can provide insight into the effect of the different planetary parameters on the PV structure. For example, decreasing the rotation rate not only decreases the planetary vorticity, but also causes a poleward shift of the jet (Fig.~\ref{fig:fig9}a), thereby contributing to the annular PV in the intermediate rotation rates (when the rotation rate is slow, the jet is too close to the pole, resulting in a monopolar PV). Alternatively, as the obliquity increases, the vortex shifts more poleward and becomes narrower, which results in an annular PV for high obliquity and fast rotation rates (Fig.~\ref{fig:fig9}).

\section{Conclusions}\label{sec:conc}
We have examined the dependence of the structure and seasonal characteristics of polar vortices in terrestrial planets on obliquity, orbital period, and rotation rate. For parameters close to Earth, the simulated vortex structure and evolution are similar to those observed on Earth (Fig.~\ref{fig:fig3}). However, when moving away from Earth's parameters, we find a wide range of vortex characteristics and seasonal dependencies. Specifically, in addition to the Earth-like regime, we identify the following regimes:
\begin{enumerate}
    \item Weak seasonality, normal climate: The jet is in the low- and mid-latitudes and has weak seasonality. This regime occurs in obliquities lower than $\sim54^{\circ}$ with short orbital periods ($\omega \leq 1/8$, left column in Fig.~\ref{fig:fig4}).
    \item Weak seasonality, reversed climate: The temperature latitudinal profile in these cases is reversed, with the maximum temperature at the poles and the minimum temperature at the equator. The flow is dominated by a wide equatorial westerly jet, with essentially no polar vortex (Fig.~\ref{fig:fig4}). This regime occurs in obliquities higher than $\sim54^{\circ}$ with short orbital periods ($\omega \leq 1/8$, right column in Fig.~\ref{fig:fig4}).
    \item Winter jet weakening: The jet experiences a weakening during midwinter. This weakening is a consequence of the winter Hadley circulation spanning from one pole to the other with air descending close to the winter pole and warming these regions adiabatically. As a result, the meridional temperature gradients are flat during midwinter, and the jet weakens. This regime occurs for slow rotation rates with high enough obliquity ($\gamma\geq30^{\circ}$, depending on the rotation rate and orbital period, left column in Fig.~\ref{fig:fig5}).
    \item Late winter jet split: The polar jet splits during late winter and spring as the Hadley circulation contracts, and becomes more hemispherically symmetric. The jet shifts equatorward and splits into its thermally- and eddy- driven components, with the thermally-driven jet located at the edge of the Hadley circulation. This regime occurs for cases with long enough orbital period (Fig.~\ref{fig:fig6}).
    \item Double jet with varied seasonality: The zonal mean zonal wind exhibits a complex seasonal cycle with periods of a double jet and a merged jet, together with a jet weakening during midwinter. In this cases, the temperature structure at the top of the atmosphere is reversed, with the warmest high-level temperatures at the pole. As a result, in addition to the jet at the edge of the Hadley cell, there is a thermally driven jet that is disconnected from the Hadley circulation during the transition seasons. During winter, as the Hadley circulation expands to higher latitudes the jets merge. Additionally, due to the high latitudes being warm at high levels, meridional temperature gradients are flat, and there is a weakening of the polar vortex during midwinter. This regime occurs at high obliquities and intermediate-slow rotation rates (Fig.~\ref{fig:fig7}).
\end{enumerate}

%Note, that some of the model assumptions do not align with Mars and Titan, for example, ocean slab as a boundary, water as a condensible. However, the model simplicity and dynamical similarity suggest that insights from our simulation are also relevant to the observed flow in these planets.

Although our model is idealized and some of the model assumptions, such as ocean slab as a boundary and water as a condensible, do not align with the climate of Mars and Titan, our results are not detached from the natural world, as similar behaviors seen in our simulations are observed in the solar system terrestrial planets. That said, the model's simplicity together with the dynamical similarity suggest that insights from our simulations relate to fundamental dynamical processes, and are relevant to the observed flow in the solar system planets. 

In addition to the Earth-like regime, which corresponds also to the polar vortex seasonal cycle on Mars \citep[e.g.,][]{waugh_martian_2016}, the winter jet weakening regime has some similarities to the seasonal cycle of Titan's polar vortex. There are increasing evidence that Titan's polar vortex exhibits sudden warming during midwinter \citep{teanby_seasonal_2019}. Additionally, Titan atmospheric models also show a midwinter polar vortex weakening \citep{lora_gcm_2015, shultis_2022}. Both the models' results and the observed warming of the vortex align well with our simulations. At slow rotation rates, during midwinter, the circulation expands, and there is a warming of polar latitudes, resulting from air descending there. This polar warming and the wide Hadley circulation flattens the meridional temperature gradient, resulting in a weaker vortex during midwinter. 

We have also found other distinct polar vortex behaviors that have no counterpart in our solar system. Some of them, such as the normal and reverse weak seasonality regimes, were previously studied in studies that neglected the seasonal cycle \citep[e.g.,][]{kaspi_atmospheric_2015, KANG2019}, and can be relevant for planets with a massive atmosphere \citep{chemke2017dynamics}, cold planets (high atmospheric radiative timescale), or ocean worlds \citep[high surface thermal inertia,][]{Hu_2017}. Such planets would experience a weak seasonal cycle. Others, such as the double jet regime, were not previously studied. A feature unique to the double jet regime is the vertical temperature structure close to the pole, where at high obliquities, there is a reversal of the meridional and vertical temperature gradients in high levels of the polar atmosphere. Given the simplified radiation scheme used in this model, there is a need to explore the sensitivity of the temperature response to different radiative parameters such as the long- and short-wave optical depths and how it is sensitive to different radiation schemes.

In addition, we have also shown that the potential vorticity (PV) meridional structure of the polar vortices can be very different from Earth's polar vortices. On Earth, the maximum PV is at the vortex's center, but we have shown that annular PV is not uncommon within the planetary parameter space explored. Across the parameter space there is high (absolute) PV poleward of the jet, with large meridional PV gradients at the jet. However, the details vary with the parameters. In particular, the gradients of PV within the vortex vary, with cases with a monotonic increase in PV towards the pole (monopolar vortex, similar to Earth's polar vortex), global PV maximum around the peak winds and a decrease towards the pole (annular vortex, similar to Mars's and Titan's polar vortex), or a local maximum of PV close to the jet and a global maximum at the pole. These variations depend on the planetary potential vorticity (i.e., rotation rate) and the vortex characteristics, mainly the vortex latitude, width, and strength, which depend on the different planetary parameters. It is important to note that similarly to \citet{scott_forcing_2020}, we found that the maintenance of an annular PV in these simulations is mainly a result of Hadley circulation transport. This suggests that although that on Mars there is a need for latent forcing to maintain an annular PV \citep{seviour_stability_2017}, this is not needed in general.

We have also shown the appearance of storm activity suppression during midwinter, when the jet is the strongest. This suppression of storms during midwinter is also observed on Earth \citep[e.g.,][]{nakamura_midwinter_1992, afargan_midwinter_2017} and on Mars \cite[e.g.,][]{LEWIS2016}. However, the suppression of storm activity on Earth, on Mars and in our simulations shows different characteristics. For example, on Earth the jet becomes more thermally driven and shifts equatorward during midwinter \citep[e.g.,][]{yuval_relation_2018}, which does not occur in our simulations. Unlike Earth and our simulations, the suppression of baroclinic activity on Mars is confined to lower levels close to the surface \citep[e.g.,][]{LEWIS2016}. Yet, the existence of a midwinter minimum in our simulations aligns with other explanations such as the jet narrowing as it strengthens \citep{harnik_effects_2004} or a threshold in the jet strength \citep{hadas_2020}. The differences in characteristics between the planets, and our simulations, need to be further studied to explore how planetary parameters and atmospheric characteristics affect the relation between storms and the polar vortex.

The fact that the polar vortices in our idealized model have a jet strength, seasonality and PV structure comparable to the observed polar vortices on terrestrial planets suggest that the dynamics observed on the solar system terrestrial planets relate to basic dynamical processes that occur in planetary atmospheres rather than specific processes that occur in a specific planet's atmosphere. This, in turn, suggests that on planets outside the solar system, there will be a large variety of climates that can also vary significantly during the seasonal cycle, depending on the planetary parameters. This variability can also substantially impact future observations from planets, and future studies need to account for it.

\begin{acknowledgments}
IG and YK acknowledge support from the Minerva Foundation and the Helen Kimmel Center of Planetary Science at the Weizmann Institute of Science. DW acknowledges support from the U.S. National Science Foundation and National Aeronautics and Space Administration.
\end{acknowledgments}

\bibliography{polar_vortex.bib}{}
\bibliographystyle{aasjournal}



\end{document}